\documentclass{webofc}
\usepackage[varg]{txfonts}   % Web of Conferences font
\usepackage{hyperref}
\usepackage{url}
\usepackage[capitalise]{cleveref}
\usepackage{array}
\newcolumntype{C}[1]{>{\centering\arraybackslash}p{#1}}
\usepackage{mciteplus}
\usepackage{ifthen} 
\newboolean{uprightparticles}
\newboolean{articletitles}
\setboolean{articletitles}{true}
\usepackage{lineno}
\hypersetup{colorlinks=true,citecolor=blue,urlcolor=blue,linkcolor=blue}
\begin{document}
\title{Translating LHCb’s Documentation}
%
% subtitle is optionnal
%
\subtitle{First experiences and ensuring maintenance}

\author{\firstname{Andy} \lastname{Morris}\inst{1}\fnsep\thanks{\email{andrew.george.morris@cern.ch}} \and
        \firstname{Zhijie} \lastname{Wang}\inst{2}\fnsep\thanks{\email{zhijie.wang@cern.ch}} \and
        \firstname{Xuhao} \lastname{Yuan}\inst{3}\fnsep\thanks{\email{xuhao.yuan@cern.ch}} \and
        \firstname{Yisheng} \lastname{Fu}\inst{3} \and
        \firstname{George} \lastname{Hallett}\inst{4} \and
        \firstname{Jingqi} \lastname{He}\inst{5} \and
        \firstname{Kai} \lastname{Liu}\inst{2} \and
        \firstname{Jiayu} \lastname{Zhao}\inst{5} \and
        \firstname{Xiaokang} \lastname{Zhou}\inst{5}
        }

\institute{
     Aix-Marseille Univ, CNRS/IN2P3, CPPM, Marseille, France
     \and
     Lanzhou University, Lanzhou, China
     \and
     Institute of High Energy Physics, Beijing, China
     \and
     Department of Physics, University of Warwick, Coventry, United Kingdom
     \and
     Central China Normal University, Wuhan, China
          }

\abstract{
        The Starterkit Lessons online and Starterkit Workshops held in Geneva each year have been the main method of onboarding newcomers to the LHCb experiment since its founding in 2015.
The new software corresponding to Upgrade 1 of the LHCb and Run~3 of datataking at the LHC has necessitated a new version of this Starterkit to be written.
This new version has made several improvements over the old one, with increased maintainability through testing of examples in CI pipelines, and has also allowed for translations of the Starterkit Lessons.
In late 2025 the Run~3 Starterkit lessons were translated into Mandarin Chinese, to allow for a sibling Starterkit event to take place in China, with a dedicated liaison role set up to ensure synchronization between the two translations.
To measure the success of the Chinese Starterkit Lessons, analytics have been added revealing that roughly 22\% of all visits to the Starterkit website will access at least one Chinese-language page.
}
\maketitle
\section{Introduction}
\label{sec:intro}

Since 2015, the main method of onboarding and training newcomers to the software used at the LHCb experiment \cite{LHCb:2008vvz} has been the Starterkit \cite{ReinsvoldHall:training, Puig:starterkit}.
This has represented simultaneously a workshop for training newcomers hosted in Switzerland each November (the Starterkit Workshop) and also two series of follow-along lessons hosted online (the Starterkit Lessons).
In general the Starterkit Workshop is taught from a mixture of the first set of Starterkit Lessons and additional material from the HSF analysis essentials lessons.

The LHCb experiment underwent an upgrade between 2018 and 2022 \cite{LHCb:2023hlw} during which most of the software also underwent major changes.
As such a new set of Starterkit lessons the ``Run~3 Starterkit'' were written with the first public version being released in January 2025.

Originally the Starterkit Workshops were held as in-person only events, with a single year of online-only during 2020 and a hybrid approach since then.
In-person attendance is still preferred to give newcomers an opportunity to meet other members of the collaboration beyond their home institute, however it can pose an issue for those who would have to travel a long way and acquire a visa to visit Switzerland: notably the case for collaborators in China.
China is relevant in this respect as it is the single country with the most LHCb collaborators: 299 as of July 2026.\footnote{The second- and third-highest being Italy with 247 and the UK with 244.}
As such it is necessary to accommodate colleagues from China who cannot easily attend the Starterkit Workshop.
The LHCb Data Processing and Analysis (DPA) project, who administrate the Starterkit Lessons, therefore decided on implementing a simultaneous `sibling event' in China to complement the European workshop.

To allow for this, the Starterkit Lessons were first translated into Mandarin Chinese with the goal of maintaining both translations simultaneously moving forward.
This was envisioned to use a mixture of automated processes and human quality assurance to ensure a high quality translation without adding a bottleneck to changes in the lessons' content.
This came with an initial process of translating all the pre-existing lessons.
At the same time, the opportunity was taken to extract all example code from the English version of the Starterkit Lessons into external scripts.
These scripts are then re-imported into the lessons from their source code, ensuring consistency across translations and also allowing their testing within a continuous integration framework.

\section{Original Translation}
\label{sec:original-translation}

% =========================================================================
% 2 Original Translation Workflow
% =========================================================================
To migrate the initial content of the LHCb Starterkit documentation into Chinese, a specialized translation workflow was developed. This workflow is designed to guarantee high semantic accuracy while addressing the time and labor costs that a purely manual translation of the set of preexisting lessons would entail. To achieve this, a hybrid translation framework was established using Large Language Models (LLMs) with human-in-the-loop validation. The system leverages a locally deployed LLM engine~\cite{11168242} to generate initial translations, which are subsequently validated and refined by native Chinese-speaking physicists to eliminate potential machine-generated hallucinations or errors.

\subsection{Model Selection and Performance Evaluation}
\label{subsec:benchmarking}

To establish the core foundation of the local deployment, benchmarking was conducted of the resource usage of three open-source LLMs available at the time: Llama-3-70B~\cite{llama3}, Qwen3.5-35B-A3B~\cite{qwen3}, and DeepSeek-R1-32B~\cite{deepseek-r1}.

All local benchmarks were executed on a standard, consumer-grade computing workstation equipped with a single NVIDIA RTX 4070 Ti Super GPU (16~GB VRAM). To accommodate the strict memory ceiling of consumer-grade hardware, models exceeding native VRAM capacity were managed via advanced model quantization (e.g., 3-bit or 4-bit GGUF/AWQ configurations) and memory streaming strategies.
The results of these hardware and throughput benchmarks are summarized in \cref{tab:model_benchmark}.

\begin{table}[htbp]
\centering
\caption{Local inference benchmark of the tested open-source LLMs.}
\label{tab:model_benchmark}
\makebox[\textwidth][c]{
    \begin{tabular}{lcC{23mm}C{23mm}}
    \hline
    \textbf{Model Name} & \textbf{Baseline Accuracy} & \textbf{Peak VRAM Footprint (GB)} & \textbf{Throughput (tokens/sec)} \\ \hline
    \texttt{Llama-3-70B}     & $\sim$80\%                & $>$16                       & $\sim$8                           \\
    \texttt{DeepSeek-R1-32B} & $\sim$80\%                & $\sim$13.2                      & $\sim$15                          \\
    \texttt{Qwen3.5-35B-A3B}      & $\sim$80\%                & $\sim$13.4                        & $\sim$28                          \\ \hline
    \end{tabular}
}
\end{table}

An initial challenge identified during this benchmark was that standard, off-the-shelf vanilla LLMs frequently failed when encountering localized high-energy physics (HEP) technical terms or LHCb-specific softwares (such as \textit{Gaudi}, \textit{DaVinci}, or \textit{TupleTool}). Quantitative manual audits indicated that the baseline technical accuracy of the vanilla translations (i.e. how often a sentence translated by the LLM would need correcting manually) hovered at only approximately 80\% across all three models. This $\sim$20\% error margin introduced severe domain inaccuracies and technical hallucinations, making raw vanilla outputs unacceptable.

The models behaved very differently with regards to runtime efficiency and memory access. The peak memory usage required by Llama-3-70B is over 16GB, forcing the engine to offload weights to slower host system memory (CPU-RAM), throttling its speed to 8~tokens/sec -- significantly slower than the alternatives. The reasoning-centric DeepSeek-R1-32B fit within the 16~GB envelope ($\sim$13.2~GB VRAM), but its intrinsic ``chain-of-thought'' loops introduced heavy computational overhead for direct translation, resulting in an intermediate throughput of 15~tokens/sec. 

The Qwen3.5-35B-A3B model achieved the most optimal operational balance. This is due to its efficient attention architecture and native compatibility with low-bit frameworks (such as FlashAttention-2~\cite{dao2023flashattention2fasterattentionbetter} via llama.cpp~\cite{kurt2026quantizationiuseunified}), it operated stably near the 16~GB VRAM ceiling ($\sim$13.4~GB) while maintaining a high throughput of $\sim$28~tokens/sec. Consequently, Qwen3.5-35B-A3B was selected as the foundational translation engine for the Starterkit translation due to its better performance.

\subsection{Retrieval-Augmented Generation (RAG) Implementation}
\label{subsec:rag_implementation}

To bridge the 20\% performance gap seen in the vanilla baseline and eliminate domain-specific hallucinations, a Retrieval-Augmented Generation (RAG) architecture was deployed, tailored specifically for technical documentation extraction~\cite{NEURIPS2020_6b493230}. By specializing the generative capabilities of the selected Qwen3.5-35B-A3B model using a verified knowledge base, the translation accuracy before human intervention was successfully elevated beyond 95\%.

The methodology used to perform the initial translation with the RAG proceeded as:
\begin{enumerate}
    \item \textbf{Document Parsing and Code Shielding:} Prior to translation, the documentation Markdown source files are parsed into logical syntactic blocks. To prevent the core language model from corrupting code syntax, embedded code snippets, code blocks, internal hyperlinks, and \LaTeX\ expressions are identified using regular expressions and automatically wrapped in placeholder token. This ensured structural isolation during the generative process.
    \item \textbf{Bilingual Terminology Indexing:} An English-to-Chinese glossary encompassing HEP experimental physics was compiled and verified by bilingual LHCb collaboration members. These term pairs were converted into high-dimensional vector embeddings using a local embedding model~\cite{reimers-gurevych-2019-sentence} and indexed inside a lightweight vector database.
    \item \textbf{Contextual RAG Retrieval:} For each semantic block extracted from the documentation, the pipeline performs a top-$K$ cosine similarity search across the vector database~\cite{8594636}. The matching bilingual pairs are dynamically retrieved to serve as localized translation anchors.
    \item \textbf{Constrained Few-Shot Prompting:} The retrieved terminology constraints are systematically injected into the LLM's system prompt using custom few-shot templates. The prompt explicitly instructs the foundation model to treat the retrieved pairs as absolute lexical invariants while reconstructing the surrounding natural language in fluent, academically appropriate Chinese~\cite{NEURIPS2020_1457c0d6}.
\end{enumerate}

Due to the strict prompt constraints imposed by the glossary indexing under the RAG framework, semantic deviations and technical hallucinations were virtually eliminated, raising the automated baseline translation accuracy across standard text blocks to over 95\%.

\subsection{Human-in-the-Loop Post-Editing and Quality Assurance}
\label{subsec:human_in_the_loop}

While the RAG-augmented Qwen3.5-35B-A3B pipeline provided a high baseline accuracy, the accuracy was not high enough to prevent misinterpretation by the reader. The final improvements to accuracy were achieved with manual review at the final stage of the translation workflow to make sure the text is correct.

Once the automated local inference concludes, the generated Chinese Markdown files are routed to a panel of bilingual LHCb experts for comprehensive post-editing and manual verification. During this phase, reviewers perform minor stylistic corrections, refine the natural flow of highly complex physics prose, and double-check that no nested code arguments were inadvertently shifted. Introducing human-assisted review steps ensures that the final technical document translation is basically accurate, free from misleading errors, and suppresses remaining mistakes caused by AI translation.
\section{Maintenance}
\label{sec:maintenance}
To ensure the longevity of the documentation, we have designed and implemented a multi-level maintenance and quality assurance system. Given the frequent software iterations and updates characteristic of High-Energy Physics (HEP) collaborations, real-time updates to tutorial documentation are essential, necessitating a robust and reliable maintenance workflow~\cite{Villanueva_2023}. This section details the supporting maintenance architecture across three dimensions: content revisions led by specialists, synchronized Chinese text updates by Chinese liaisons, and automated Continuous Integration (CI) testing workflows.

\subsection{Expert Auditing}
Ensuring the accuracy of tutorial materials is important, as examples that don't work ``out of the box'' or misconceptions regarding core physical concepts can severely impede the analytical workflows of early-career researchers. To reduce these risks and ensure the reliability of the content, a feedback-driven optimization mechanism is used.

This begins with a standardized peer-review process in which dedicated LHCb experts conduct regular quality checks on technical documentation.
The feedback that comes from the review reports on areas where the Starterkit could be improved, and makes suggestions to aid the utility from both graduate students and senior physicists based on their day-to-day research.
Upon receiving this feedback, the expert panel meets to evaluate the input and adjust the documentation as needed. Approved optimization plans are directly translated into targeted revisions, along with the changes to the original English text. This forms a sustainable operational system that enables continuous iteration and refinement of the collaboration's documentation, ensuring the Starterkit is as pedagogical as possible while maintaining high quality over the long term~\cite{Albrecht2019roadmap}.

\subsection{Chinese Liaison Synchronization Update}
The synchronous update of Chinese documents is the responsibility of a dedicated Chinese liaison attached to Work Package 7 (Training and Documentation) of the LHCb DPA project. The liaison's focus is entirely on translating, validating, and applying upstream modifications to guarantee that the Chinese codebase remains accurate and up-to-date with the English documentation.

The operational pipeline of the liaison framework is structured across the following core synchronized workflows:
\begin{enumerate}
    \item \textbf{Merge Request Triggered Sync:} When a merge request is submitted, reviewers will promptly notify the Chinese liaisons to translate the revised text. This workflow ensures that by the time the code or content is merged, the Chinese documentation is already fully up-to-date.
    \item \textbf{AI-Assisted Translation Workflow:} The liaisons undertake a two-stage process to integrate updates. First, a fine-tuned LLM generates the initial translation of the modified text; subsequently, liaisons perform human verification to refine the output~\cite{chen2024learning}. This approach guarantees both rapid deployment and the accuracy of the Chinese documentation.
\end{enumerate}
% By prioritizing the core mission of source-to-target alignment, this dedicated liaison structure guarantees that the active Chinese documentation is updated promptly without altering the upstream English ecosystem or interfering with user-feedback triage.

\subsection{Automated CI Testing Framework}
To enforce rigid computational quality standards once both the reviewers and Chinese liaisons have verified the modifications to be error-free, the repository leverages an automated Continuous Integration (CI) testing pipeline built via GitLab runners~\cite{refId0}. The automated testing framework executes deterministically upon every approved update, focusing on two elements of quality assurance:
\begin{itemize}
    % \item \textbf{Python Code Verification:} Because high-energy physics starter guides are dense with inline executable blocks and standalone configuration scripts---such as Python options for the Gaudi architecture or initialization scripts for DaVinci analyzers---any broken syntax renders the tutorials non-functional. The CI pipeline extracts these Python scripts from the documentation source and executes them in an isolated, synchronized container environment, verifying that all imports, API parameters, and data-loading macros run perfectly without generating runtime exceptions~\cite{Migliorini2020pipelines}. % TODO Rewrite entirely
    \item \textbf{Python Code Verification:} To ensure that code examples work and are consistent across translations, almost all code blocks are separated into their own source files.
    These are then reinjected into the lessons across translations, ensuring that they cannot be different.
    These examples are then separately tested to verify that they execute and provide the expected results.
    \item \textbf{Document Formatting and Link Validation:} Beyond functional code, the CI pipeline enforces structural syntax checks on the overall text files, using automated linters~\cite{7962384}. Simultaneously, the testing runner deploys an automated link checker that verifies the availability of every hyperlink embedded within the modified documentation files, flaggin dead external links, HTTP error responses, or shifted interior anchor hashes.
\end{itemize}
A successful pass through the entire CI pipeline is required before a modification can be merged into the Starterkit. Additionally, an extra optional step is available to publish a preview version of the site with the proposed changes, allowing for manual review e.g. of newly added lessons. By integrating human auditing with this automated gating mechanism, the system guarantees that the final deployment is understandable to newcomers and that the code examples will run, ensuring the rigor of the documentation.
\section{Uptake}
\label{sec:uptake}
\begin{figure}
    \centering
    \includegraphics[width=0.8\linewidth]{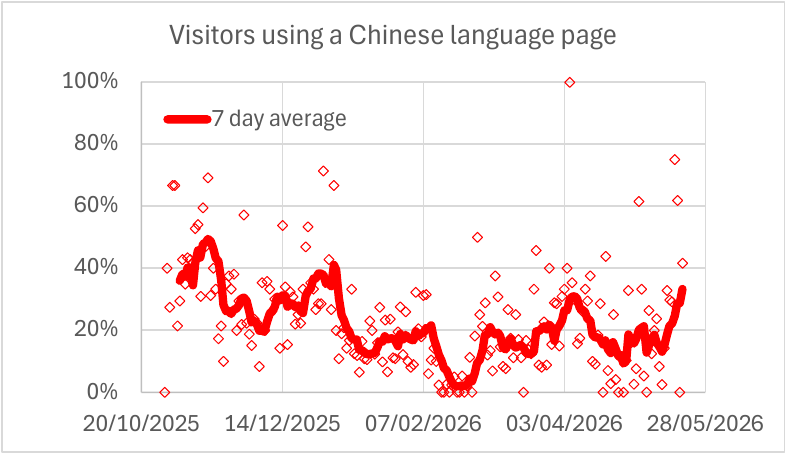}
    \caption{The proportion of visits accessing a Chinese-language page on the Starterkit website and a rolling average of each day with the previous six accounting for daily variations in number of visits.}
    \label{fig:zh_ratio}
\end{figure}

To measure the popularity of the translation, basic analytics were added to the Starterkit website using Matomo via CERN's IT group. This provides tracking of the pages users visit without saving any personal information about the users (in compliance with GDPR laws where applicable).
Further safeguards are added to the analytics to ensure that neither local builds of the site nor previews of merge requests are included in the data. Information is stored for 500 days tracking `visits' of users.
A `visit' here represents a person connecting to the website, and may include visiting multiple different pages before disconnecting or the session timing out, only when reconnecting will it be counted as a separate visit \cite{matomo_visit}.
Using information about which pages were viewed in a visit allowed the tracking of the proportion of users who visit a Chinese-language page.
This is plotted in \cref{fig:zh_ratio} showing the ratio for each day, as well as a rolling average showing the ratio for a day and the previous six. This average is done accounting for variations in the number of visits each day.
What was found after the initial surge of interest was a fairly stable usage of the Starterkit with, on average, 22\% of visits viewing a Chinese-language page. A noticeable dip is also noted corresponding to lunar new year.

The addition of analytics has also allowed for tracking the popularity of the Run~3 Starterkit: receiving an average of $30\pm 1$ visits a day between 29/Oct/2025 and 19/May/2026 with an average of $6.7 \pm 0.4$ visits accessing at least one Chinese-language page.
\section{Conclusions}
\label{sec:conclusions}

Since its inception, the Starterkit Lessons and Starterkit Workshop have been key to the LHCb experiment's onboarding allowing newcomers to get to grips with the collaboration's software: this has not changed in Run~3.
The Run~3 Starterkit Lessons, including their Chinese translations, have proved very popular with tens of visits every day.
This popularity is due to its accessibility, with lessons being written with a pedagogical approach aimed at newcomers but also due to the unit tests ensuring its maintainability -- the examples have been tested to work with each iteration and are updated when necessary.
A system has been established where updates to a page in one language will be translated by a specialized AI to the other which is then edited by a dedicated liaison; allowing the pages to be synchronized efficiently while still ensuring a human has checked all pages in all translations before they go live.
This efficiency will be especially key each year in November, as the Starterkit Workshops typically lead to the highest density of changes made to the Starterkit Lessons ahead of the workshop, keeping synchronicity during this time will be paramount.

With these systems in place, the groundwork is in place for the first sibling Starterkit Workshops to begin from this year in China, being taught in Chinese, to complement the European workshop in November 2026. It should also be noted that these systems are extensible, should other places wish to have their own sibling events: there are a significant number of LHCb collaborators in Brazil for example who could organize their own dedicated LHCb onboarding events if interest arose.

\section*{Acknowledgements}

The authors wish to thank Jo\"el Closier and Subhashis Suara for their help in setting up the analytics with CERN IT.

The authors would also like to thank the LHCb DPA project, in particular Nicole Skidmore and Chris Burr for their help in setting up the Training and Documentation project which created the Run~3 Starterkit.
\bibliographystyle{LHCb}
\bibliography{references}
\end{document}